# Au nanowire junction breakup through surface atom diffusion

**Simon Vigonski[1,2], Ville Jansson[2], Sergei Vlassov[3], Boris Polyakov[4], Ekaterina Baibuz[2], Sven Oras[3], Alvo Aabloo[1], Flyura Djurabekova[2] and Vahur Zadin[1,2,*]**

[1]Institute of Technology, University of Tartu, Nooruse 1, 50411, Tartu, Estonia

[2]Helsinki Institute of Physics and Department of Physics, P.O. Box 43 (Pehr Kalms gata 2), FI-00014 University of Helsinki, Finland

[3]Institute of Physics, University of Tartu, W. Ostwaldi 1, 50411, Tartu, Estonia

[4]Institute of Solid State Physics, University of Latvia, Kengaraga 8, LV-1063, Riga, Latvia

*Corresponding author email: vahur.zadin@ut.ee

**Abstract.** Metallic nanowires are known to break into shorter fragments due to the Rayleigh instability mechanism. This process is strongly accelerated at elevated temperatures and can completely hinder the functioning of nanowire-based devices like e.g. transparent conductive and flexible coatings. At the same time, arranged gold nanodots have important applications in electrochemical sensors. In this paper we perform a series of annealing experiments of gold and silver nanowires and nanowire junctions at fixed temperatures 473, 673, 873 and 973 K (200, 400, 600 and 700 °C) during a time period of 10 minutes. We show that nanowires are especially prone to fragmentation around junctions and crossing points even at comparatively low temperatures. The fragmentation process is highly temperature dependent and the junction region breaks up at a lower temperature than a single nanowire. We develop a gold parametrization for Kinetic Monte Carlo simulations and demonstrate the surface diffusion origin of the nanowire junction fragmentation. We show that nanowire fragmentation starts at the junctions with high reliability and propose that aligning nanowires in a regular grid could be used as a technique for fabricating arrays of nanodots.

## 1. Introduction

Gold nanostructures are of considerable interest for their optical, mechanical and electrical properties. For example, gold nanopillar arrays have been used as highly efficient electrodes for detecting bioelectrical signals [1,2], where their aspect ratio has proven to be of great importance. Plasmonic trapping of colloidal particles has been demonstrated using a gold nanopillar [3]. Other structures with a high surface to volume ratio are porous gold films [4] and nanoparticles [5], which are used in catalysis and electrochemical sensors. In particular, ordered arrays of gold nanoparticles can be used for biochemical sensing [6], as well as wavelength-specific photodetectors [7]. Precise fabrication techniques allow for a significant level of control of the resulting geometries and properties [8,9]. Production of nanodot arrays is currently performed using vacuum evaporation [10] or using a polystyrene template [11].

Metal, including gold, nanowires are a rapidly expanding area of research as well. For example, gold nanoparticles are used as a catalyst in the production of nanowires for solar cell applications [12]. Gold nanowires themselves can be used in transparent electrodes for flexible displays [13]. A particularly important point relating to electrodes is the stability of nanowires under thermal loading – surface energy minimization driven by thermally activated diffusion leads to breakup of nanowires. This has been observed for Ag [14], Cu [15], as well as Au [16].

The behavior of nanostructures at elevated temperatures can differ drastically from the macroscopic situation. It is well known that small nanoparticles melt at a significantly lower temperature compared to bulk, and the melting temperature depends on size [17]. Moreover, if we consider the time factor, then the situation becomes even more complicated. Given enough time for surface diffusion processes to happen, it is possible to observe drastic changes in nanostructure morphology at very moderate temperatures. For instance, an Au nanoparticle of 5 nm in diameter starts to melt at approximately 1100 K (830 °C) [18], and for particles over 10 nm melting temperatures are comparable to bulk values (1337 K / 1064 °C). When we consider processes like surface diffusion and Rayleigh instability [16], we can see fusion and fragmentation of Au nanostructures at temperatures as low as 473 K (200 °C) [19]. This phenomenon is of great importance in all applications where nanowires are exposed to elevated temperatures or require thermal treatment before use to remove surfactant and other organic residuals. In particular, for proper functioning of nanowires-based transparent conductive coatings [20], continuous pathways are absolutely essential for electrons to ensure sufficient electrical conductivity. Heat-induced fragmentation of nanowires will prevent functioning of nanowire-based electronics.

Using atomistic computer simulations provides insights into the microscopic processes driving nanoparticle evolution which are difficult to observe experimentally. Molecular dynamics (MD) simulations have been used extensively to study the elasticity and plasticity of gold nanowires (e.g. [21]) and nanopillars (e.g. [22]), where the significant role of surface stress has been determined. Pareira and Silva [23] simulated a cold welding process of gold and silver nanowires with MD, where they identified diffusion, surface relaxation and reconstruction as the main mechanisms of interest. Monte Carlo (MC) methods are used to simulate longer time periods than those approachable by MD simulations. For example, Kolosov *et al.* [24] studied the coalescence of gold and copper nanoparticles. The Kinetic Monte Carlo (KMC) method was used by He *et al.* [25] to simulate structural transitions in gold nanoparticles. Müller *et al.* [26] showed the formation and breakup of a Ge nanowire using lattice KMC simulations.

In this paper we examine the breakup of Au nanowire junctions under thermal treatment and develop a gold parametrization for the Kinetic Monte Carlo code Kimocs [27] in order to simulate the breakup process. Kimocs is specially designed to simulate atomistic diffusion processes on metal surfaces. It was initially developed for copper, but has also been successfully applied for Fe nanoparticle simulations [28], where it was demonstrated that certain combinations of temperature and deposition rate result in cubic nanoparticle shapes. Kimocs requires that the transition energy barriers for all possible surface processes are known in advance.

We show that the thermally activated diffusion of surface atoms results in preferential breakup at the nanowire junction. Based on the experimental and simulation analysis we suggest a method for manufacturing periodic, well controlled arrays of nanodots.

## 2. Materials and Methods

### 2.1. Experimental setup

The experimental part of this work was done using gold and silver nanowires. Silver nanowires were purchased from Blue Nano (USA), while gold nanowires were prepared by us as described below.

#### 2.1.1. Synthesis of nanowires

The Au nanowires used in the current study were synthesized using a 3-stage process according to a technique described in detail in [29]. First, a seed solution of Au nanoparticles was prepared with 18 ml of 0.025 M sodium citrate and 0.1-0.2 ml of 0.0005 M $HAuCl_4$ solution added into a 25 ml glass bottle. Ice cold solution of 0.01M $NaBH_4$ was separately prepared. A volume of 0.6 ml of the $NaBH_4$ solution was added into the solution of sodium citrate with gold precursor while stirring vigorously. The resulting seed solution (SS), slightly orange in color, was used for synthesis of Au nanowires within 10 min after preparation.

Next, a growth solution (GS) was prepared in a 300 ml vessel by mixing 238.5 ml of 0.2 M hexadecyltrimethylammonium bromide (CTAB) and 10 ml of 0.0001 M of $HAuCl_4$. The solution had intensive yellow color. Next, 1.5 ml of 0.1 M ascorbic acid was added, making the solution colorless. The freshly prepared growth solution was divided into two 25 ml glass bottles labeled A and B, and 200 ml in vessel C. A volume of 0.25 ml of concentrated $HNO_3$ was added into vessel C. An amount of 200 μl of the gold seed solution (SS) was added into bottle A and stirred for few seconds (the solution color was pink). Then 200 μl of solution in bottle A was transferred to bottle B and stirred for several seconds (the solution color was crimson-violet). Finally, 100 μl of solution in bottle B was transferred to vessel C and mixed for several seconds (the solution was colorless in the beginning, but became slightly orange-brick color after 1h). The solution was kept in at 25 °C for 12 hours. Precipitates of gold nanowires can be observed on the bottom of the tube after the reaction. The supernatant was poured out, and the precipitation was re-dispersed in 5 ml deionized water. Remaining CTAB allows storing the Au NWs suspension at least for one year.

#### 2.1.2. Preparation and experimental analysis of samples

The solution with nanowires contained a high amount of surfactant (hexadecyltrimethylammonium bromide). In order to reduce the amount of the surfactant, a special procedure was performed. The solution containing the nanowires was left intact for several hours until all nanowires settled out and the solution became transparent. Immediately prior to the preparation of samples the liquid above the precipitate was removed and replaced by distilled water. The new solution was stirred until formation of a uniform mixture and then transferred to separate Si wafers by drop-casting. In total, five samples were prepared.

A series of annealing experiments were performed at fixed temperatures 473, 673, 873 and 973 K (200, 400, 600 and 700 °C) during a time period of 10 minutes. However, only one temperature was used for each sample. In addition, one sample was treated at 973 K (700 °C) for 1 minute. The procedure of thermal treatment consisted of heating the furnace up to the required temperature and then inserting the sample for the chosen period of time. Thermal treatment was performed in air atmosphere.

Micrographs of nanowires before and after thermal treatment were obtained with high-resolution scanning electron microscope (HR-SEM, Helios Nanolab 600, FEI) and transmission electron microscope (TEM, Tecnai GF20, FEI).

### 2.2. KMC model development

For simulating the Au nanowires we use the Kinetic Monte Carlo for Surfaces code (Kimocs) [27]. Kimocs is an atomistic Kinetic Monte Carlo code for simulating single crystal structures. Kimocs is based on a rigid lattice where atoms can occupy well-defined lattice sites. A transition occurs when an atom jumps from an occupied lattice site to a neighboring vacant lattice site with a rate given by an Arrhenius type equation:

$$\Gamma = \nu \exp\left(-\frac{E_m}{k_B T}\right) \tag{1}$$

where $\nu$ is the attempt frequency, $E_m$ is the migration energy barrier for the transition, $T$ is the temperature and $k_B$ is the Boltzmann constant.

To conduct a simulation, the attempt frequency and migration barriers for all possible transitions must be known in advance. Different transitions are characterized by the number of first and second nearest neighbors of the jumping atom in the initial and final positions (see [27] for details). For simplicity we do not take into account the positioning of the neighbors, only their number, thus drastically reducing the number of possible transitions. For each such transition, the migration barrier is calculated using an automated tethered NEB process (see section 2.2.2 below). As a further simplification, the attempt frequency is taken to be equal for all transitions and calculated by fitting nanopillar relaxation times to molecular dynamics results (section 2.2.3 below).

As a result, atom jumps are characterized by migration barriers, which are calculated in molecular dynamics from the interactions of the jumping atom with the local atomic environment. Any atom can move to any neighboring vacant lattice site by overcoming the energy barrier. Although only first nearest neighbor jumps are included in the current work, other transitions are in principle possible as a sequence of jumps that may include intermediate metastable positions. In this way, an adatom can, for example, cross the edge between two different surfaces.

#### 2.2.1. Potential selection

In order to adjust the model for use with Au, a full parametrization of an Au potential had to be made. We selected the Embedded Atom Method (EAM) potential by Grochola *et al.* [30]. The energy barriers calculated for our model are based on an interatomic potential, which describes the interactions between all the atoms in the system. The choice of the potential is of utmost importance as it determines the values for all the energy barriers, which in turn dictate the evolution of the system. Interatomic potentials are typically fitted to specific experimental or *ab initio* parameters of interest. For our purposes, the most important properties are the surface energies, specifically the ordering between the energies of {111}, {110} and {100} surfaces. According to *ab initio* calculations by Vitos *et al.* [31], the surface energies for gold are (in increasing order; J/m$^2$): {111} – 1.283, {100} – 1.627, {110} – 1.700. Thus, the {111} surface is the most stable and the {110} surface least stable of the ones mentioned. Although the surface energies reported by Grochola *et al.* using their potential (J/m$^2$; {111} – 1.197, {100} – 1.296, {110} – 1.533) are somewhat lower than the *ab initio* values, as well as the experimental average surface energy

(1.851 J/m² at 25 °C and decreasing with increasing temperature [32]), it has the closest surface energy values to the ab-initio or experimental results, while maintaining correct surface stability order.

### 2.2.2. Migration barrier calculations

Transition processes in Kimocs are defined by the number of first and second nearest neighbors of the jumping atom in the initial and final positions. Although the specific positions of the neighbors are not taken into account during KMC simulations, a single specific neighborhood, defined as a permutation, must still be selected when calculating migration energies. Several different permutations correspond to the same Kimocs process. For each process, we look at all possible permutations and choose the one with the lowest sum of initial and final position energies to use for calculating the migration barrier.

After a permutation has been selected, we proceed with migration barrier calculations using the Nudged Elastic Band (NEB) method [33]. The spring constant for NEB was 1 eV/Å². In addition, a tethering approach was used, where background atoms are tethered to their initial positions in each NEB image using an additional tethering spring constant 2 eV/Å². This greatly improves the stability of the system in case of processes with few neighbors. As a result, almost all of the possible processes can be calculated in this way. For processes that remain unstable despite the tethering, the formula for spontaneous processes from [27] is used.

The details of permutation selection and tethered NEB migration barrier calculations are more fully presented in [34].

### 2.2.3. Attempt frequency calculations

The physical time for each KMC step is calculated based on the sum of transition rates of all possible processes at that time [27]:

$$\Delta t = \frac{-\log u}{\sum_i \Gamma_i} \quad (2)$$

where $\Gamma_i$ is the rate for a single process calculated using eq. (1) and $u \in (0, 1]$ is a uniform random number.

Since $\nu$ is taken to be equal for all processes, it can be taken out of the summation. Thus, the total time for a process to occur is $t = N \cdot \langle \Delta t \rangle$, where $N$ is the number of steps and $\langle \Delta t \rangle$ is the average time interval for a single step. Taking into account eq. (1), and assuming that the number of possible processes at each step and their migration barriers do not vary considerably over the whole simulation, the expression for the total time is

$$t = \frac{1}{\nu} c \cdot \exp\left(\frac{E}{k_B T}\right) \quad (3)$$

where $E$ is the average effective transition energy barrier and $c$ is a factor which incorporates the average number of possible transitions in the system and is proportional to the number of simulation steps.

As a result, simulations can be conducted using the value 1 for the attempt frequency, leading to results in normalized time units. The total normalized time can later be divided by the fitted attempt frequency to transform it into physical time.

To estimate the attempt frequency, we use the same approach as detailed in [27]. We fit the relaxation time of a nanopillar on the {110} surface to MD results. A nanopillar with a rectangular cross-section (dimensions 2.0 × 2.8 × 1.7 nm; 12 monolayers, see [27] for details) is relaxed in both MD (using LAMMPS [35]) and KMC. The time taken for the pillar to reach half its original height is recorded from the MD simulations and compared with the normalized time for the same process to occur in KMC. The attempt frequency is then calculated from the ratio of these two times.

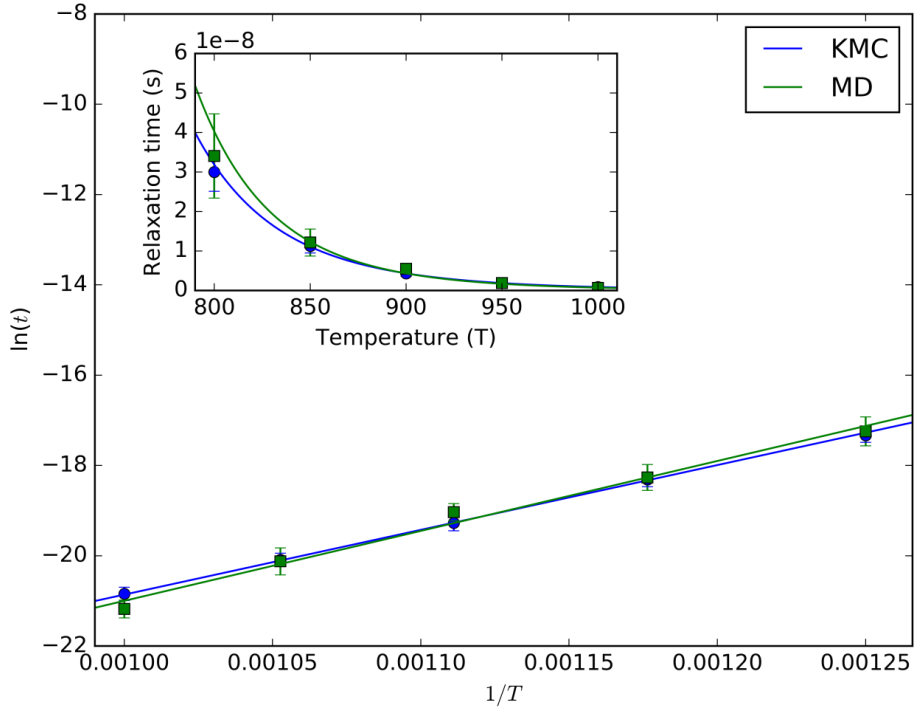

**Figure 1. Relaxation time for a nanopillar in MD and KMC simulations depending on temperature for the temperatures 800 K, 850 K, 900 K, 950 K, 1000 K. KMC relaxation time has been normalized to the MD time to minimize discrepancies. Insert: non-linearized version of the same data.**

Figure 1 shows the relaxation time of the pillar for MD and KMC simulations at different temperatures. For each temperature, the system relaxation was performed for 10 cases with different random seeds to obtain a statistical estimate. Taking into account eq. (3), the graph has been linearized by plotting the logarithm of the relaxation time against the inverse of the temperature. The different slopes indicate that the average effective transition energy barrier (parameter $E$ in eq. (3)) differs between MD and KMC. This is not surprising, since the method of calculating migration barriers makes several assumptions and simplifications (e.g. the rigid lattice and only nearest-neighbor jumps).

The intercepts of the linear fits depend on the attempt frequency. Because of the difference in slopes between MD and KMC, the relaxation times cannot be made equal for all temperatures simultaneously. We selected an attempt frequency value that minimizes the sum of the differences between the measurement points. The resulting value is $\nu = 1.22 \times 10^{17} \text{ s}^{-1}$ and it has been used to normalize the KMC data points. The fitted lines intersect at 895 K.

The difference from typical MD atom oscillation frequency of ~$10^{13}$ s$^{-1}$ could possibly be explained by the fact that second nearest neighbor or longer jumps are not included in the model. These long jumps are known to have attempt frequencies that can be even four to seven orders of magnitude higher than the nearest neighbor jumps in the case of tungsten [36]. However, in our model, the same surface evolution is achieved by only allowing nearest neighbor jumps while second nearest neighbor jumps etc. are covered by almost immediate follow-up jumps from unstable locations to stable ones. Since a series of such short jumps may happen with a different probability compared to a single long jump, the high attempt frequency we obtain by comparing with MD might be an indication that the long jumps may play a role. The sum of transition rates in Eq. 2 does not include these long jumps, increasing the time taken at each KMC step. However, since our model gives good agreement with experiments for the surface evolution (see section 4), the inclusion of long jumps would likely only affect the time estimate and thus lower the attempt frequency. The attempt frequency used in the current model can be seen as a normalization factor to fit the timescale to the more accurate MD method.

Similarly to previous results of attempt frequency calculations for Cu [27], the general temperature dependence using the two methods is similar. KMC tends to underestimate the relaxation time at lower temperatures and slightly overestimate it at higher temperatures. The variance in both MD and KMC increases significantly with decreasing temperature. The fact that the variance between repeated runs is the same for both MD and KMC, shows that it is caused by the underlying energetics of the system, rather than the method used to calculate the relaxation process.

The set of calculated migration barriers is included as supplementary material and is available online at http://iopscience.iop.org/article/10.1088/1361-6528/aa9a1b/data.

## 3. Results

### 3.1. Experimental results

According to SEM observation of an untreated sample (Figure 2), synthesis yielded uniform high-aspect ratio nanowires with regular shape, well-pronounced facets [inset in Figure 2 (a)] and smooth surface, indicating a crystalline structure, confirmed by TEM imaging [Figure 2 (b) and (c)]. Based on the SEM and TEM images, as well as according to the literature data [37,38], obtained nanowires were grown along <110> direction and had pentagonal structure with outer planes being {100}. In addition to nanowires, the mixture contained nanorods, nanoparticles and plates.

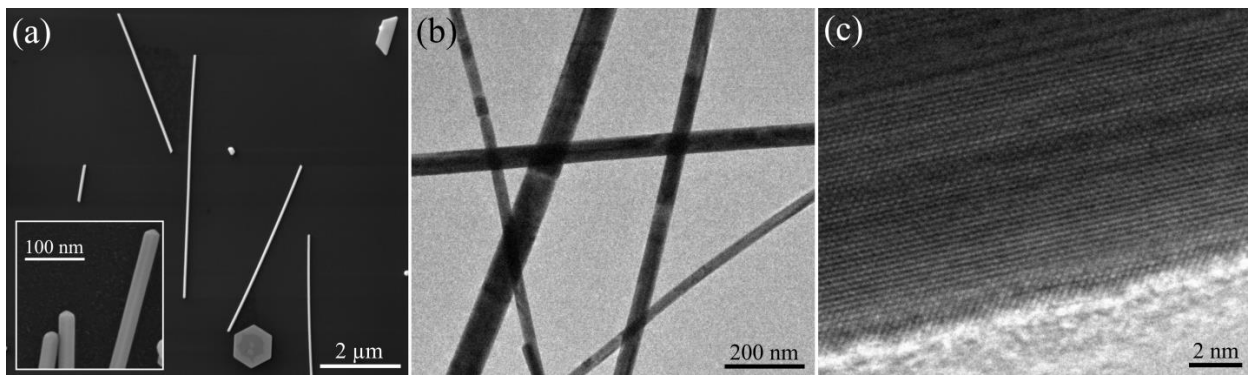

**Figure 2. SEM (a) and TEM (b, c) images of untreated Au nanowires.**

After treatment at 473 K (200 °C) for 10 minutes, most of the separate nanowires did not show any noticeable signs of changes in morphology and unity. However, decomposition (fragmentation) of nanowires was observed at places where they cross or contact [Figure 3 (a)]. Initial positions of nanowires can be deduced from traces left on the substrate by the surfactant. It can be seen that Au atoms migrated towards the contact point causing decomposition of nanowire ends.

At 673 K (400 °C) the phenomenon known as Rayleigh instability [39] appeared. Namely, in addition to decomposition at crossing and contact points [inset in Figure 3 (b)], some nanowires appeared to be fragmented to shorter pieces [Figure 3 (b)]. Fragments had the same regular faceted structure as original nanowires.

At 873 K (600 °C) a large fraction of the nanowires was fragmented and fragments were shorter, although still had regular faceted structure [Figure 3 (c)].

At 973 K (700 °C) both for 1 min and 10 minutes most of the nanowires were fragmented to faceted nanoparticles [Figure 3 (d)].

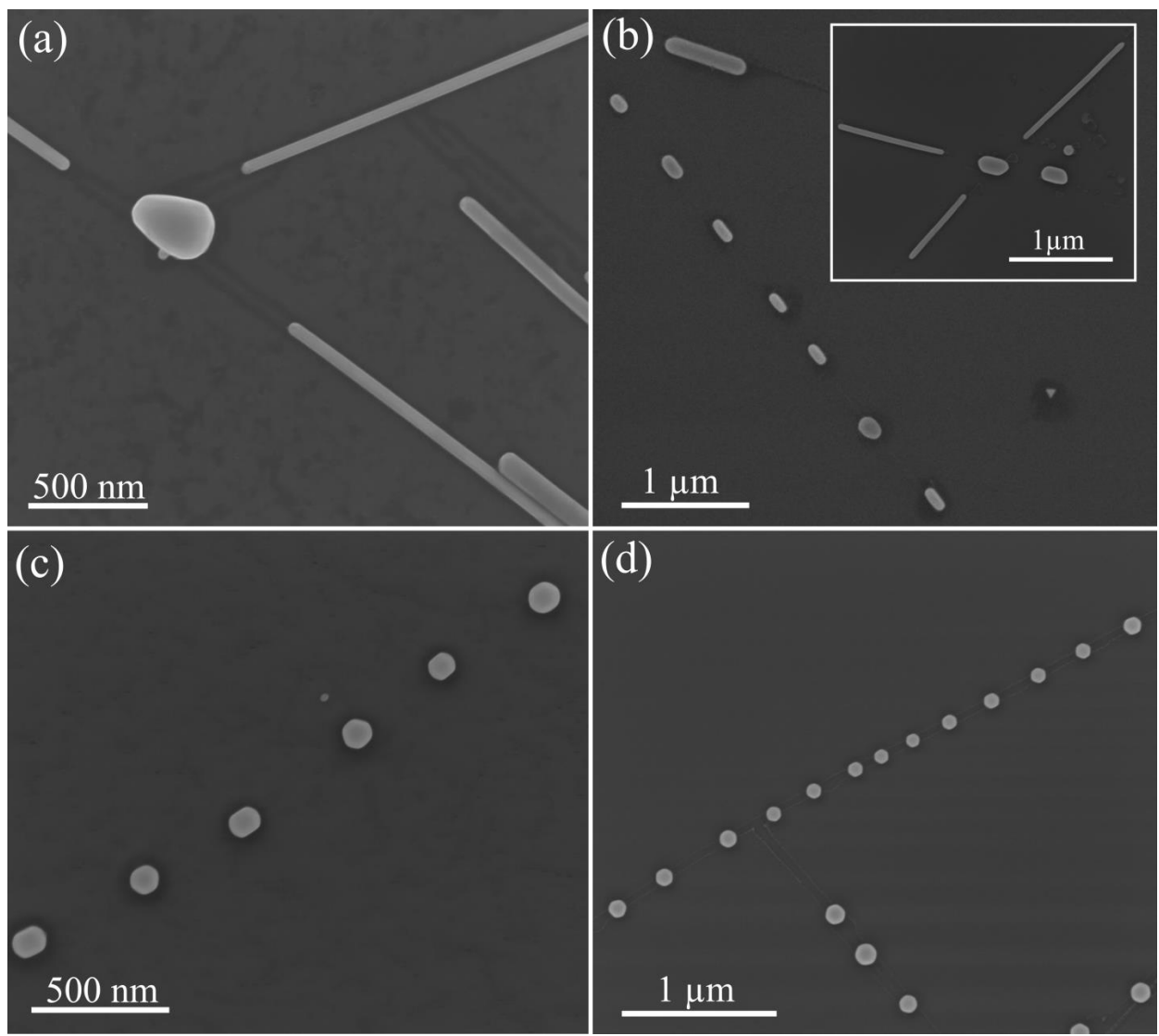

**Figure 3. SEM images of Au nanowires after thermal treatment for 10 minutes at 200 °C (a), 400 °C (b), 600 °C (c) and 700 °C (d).**

It should be noted, that even at 973 K (700 °C) some intact nanowires or long nanowire fragments were found indicating that thermal stability and onset of fragmentation process may be very sensitive to the presence of certain critical defects in nanowires.

Similar annealing experiments were performed also on Ag nanowires and it was found that silver is even less stable at mild heating. Already after 10 minutes at 398 K (125 °C), a considerable fraction of the nanowires were broken at crossing points (Figure 4). Note, that in Figure 4 (b) the intermediate state of fragmentation at the crossing point is pictured. It can be seen that material starts to diffuse from one nanowire to another.

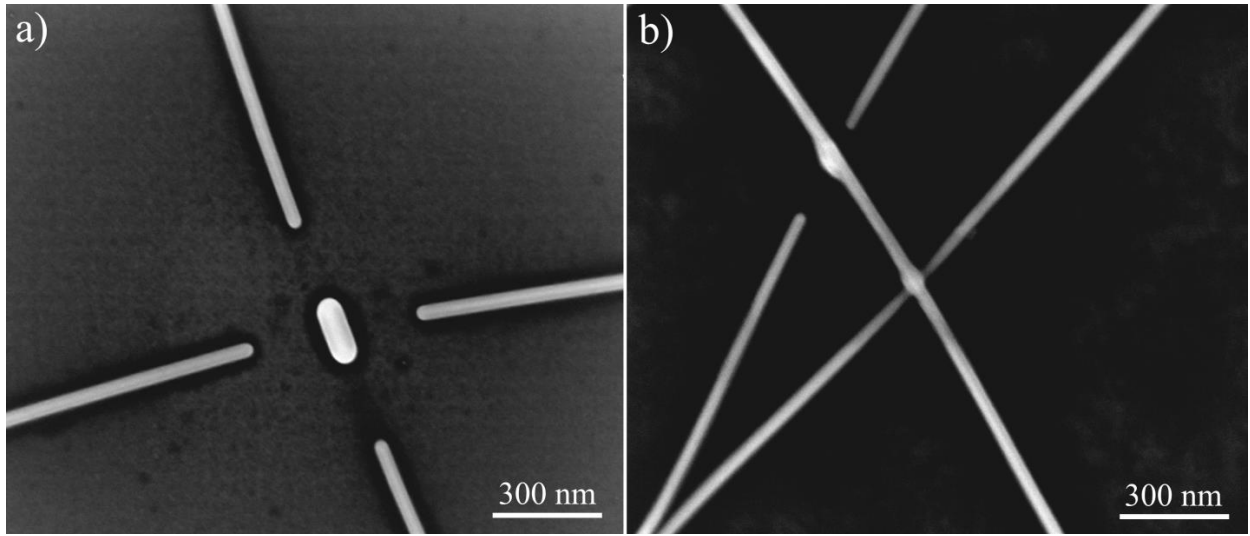

**Figure 4. Fragmentation of Ag nanowires at crossing points as a result of thermal treatment for 10 minutes at 398 K (125 °C).**

We would like to note, that we made a series of high-magnification SEM images of crossing nanowires before heating to compare the same nanowires before and after thermal treatment. However, it was found that all crossing nanowires that were previously exposed to focused electron beam (e-beam) irradiation, survived thermal treatment without any noticeable signs of morphological changes. At the same time surrounding nanowires were fragmented and broken.We believe that this effect is caused by electron beam induced carbon deposition caused by the presence of surfactant (carbon containing organics) on the surface of nanowires and substrate before heating. This phenomenon is well known in the field of electron microscopy and is called "electron beam induced deposition" (EBID).[40–42]. Organic molecules decompose under focused e-beam and re-deposit on the surface of the nanowires forming a dense carbon coating. This coating, for instance, may hinder the mobility of atoms and prevent fusion and fragmentation of nanowires.

## 3.2. KMC simulation results

### 3.2.1. Rayleigh instability of a nanowire

The Rayleigh instability driven breakup of nanowires was simulated using the developed KMC model in order to validate it. This process is driven by surface energy minimization, where the resulting nanoclusters tend to be bounded by {111} surfaces.

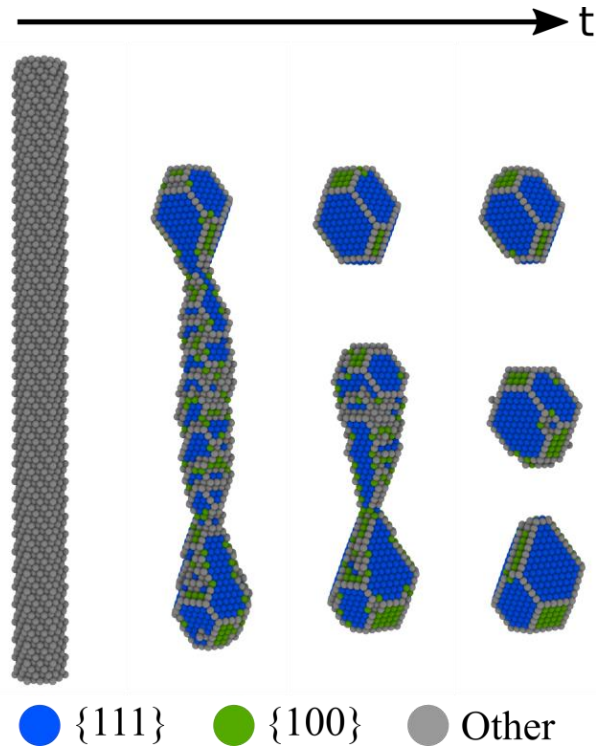

**Figure 5. Breakup of <111> gold nanowire with radius 1 nm into clusters due to Rayleigh instability at 1000 K. The system is periodic along the wire. Atoms are colored according to surface type.**

Figure 5 shows four snapshots of a <111> nanowire (the <111> crystal direction was along the wire) as the breakup progresses. The nanowire's initial radius is 1 nm and the simulations are run at 1000 K. Atoms are colored according to type of surface they belong to. The surface type is determined by inspecting the number of nearest neighbors. The wire surfaces are initially {110}, which transition into the more energetically favorable surfaces {111} and {100}. As a result, the nanowire breaks up into three nanoclusters.

To obtain sufficient statistics of the resulting nanoparticle size and separation, we used smaller nanowires with a radius of 0.5 nm. For surface diffusion driven nanowire breakup, the average nanoparticle diameter and separation are related to the initial nanowire radius. For a 0.5 nm nanowire, the theoretical average particle diameter is $d = 1.89$ nm and the average separation is $\lambda = 4.45$ nm [16,39,43]. From a series of simulations with <100> and <111> nanowires, we observed the formation of a total of 210 clusters. <110> wires are much more stable when it comes to surface diffusion processes and they do not break up in a reasonable simulation timeframe. The average measures for the 210 observed particles in our simulations are: $d = 2.01 \pm 0.17$ nm and $\lambda = 4.92 \pm 1.07$ nm. These results correspond very well with theoretical predictions, as well as other simulations [26], and confirm the validity of our parametrization.

### 3.2.2. Nanowire junctions

The same surface diffusion mechanism that is responsible for the breakup of a single nanowire acts when two wires are touching. Figure 6 depicts a sequence of simulation snapshots as a nanowire junction undergoes breakup. The simulation box is periodic along the wires, so both wires can be thought of as being infinitely long. Both wires are <100> in this case. The nanowire radius was 1 nm and the simulations were performed at 1000 K.

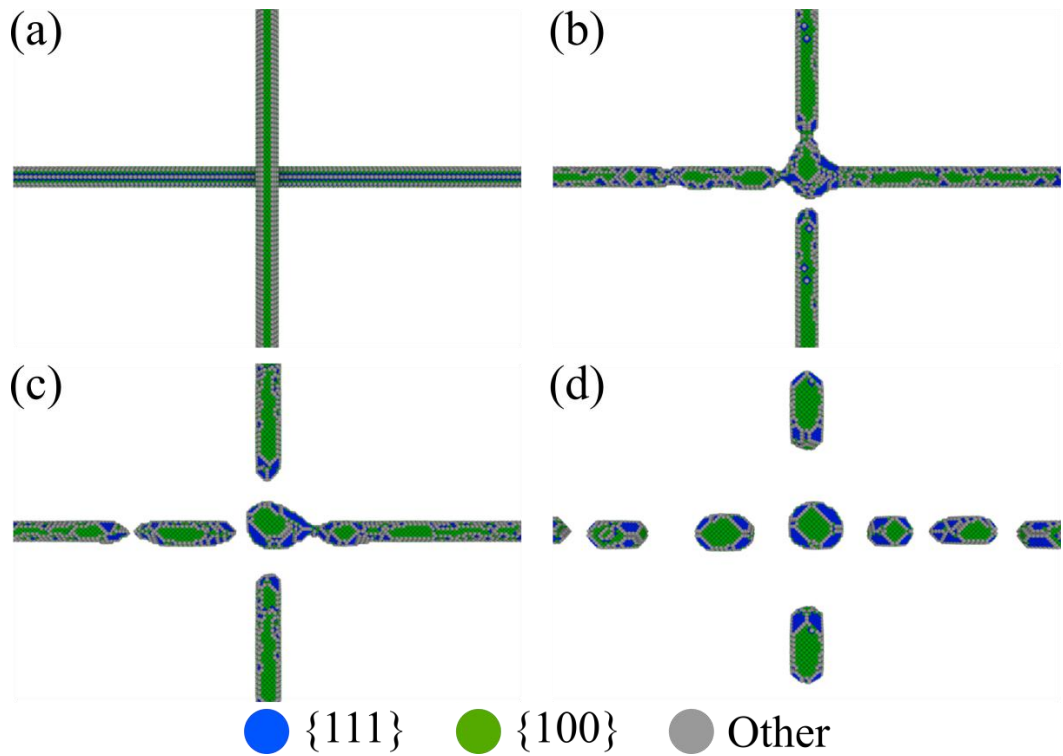

Figure 6. Breakup of a 1 nm radius nanowire junction where the crossing wires lie on top of each other at 1000 K (a). Atoms start collecting in the junction region (b), leading to a separation of the central droplet (c). Eventually, wires decompose into droplets (d).

The simulation was repeated 20 times and the time of the first detachment was recorded (the moment in Figure 6 (b)). The average time for a first detachment to occur was $4.0 \pm 0.8$ ns. In all cases the first detachment happened near the junction and almost always the central cluster was the first to form completely, although in some cases a nearby cluster would form before the central cluster could detach from all four sides. A single longer simulation was run at 800 K where the time until the first detachment was 140 ns (see movie 1 in supplementary materials).

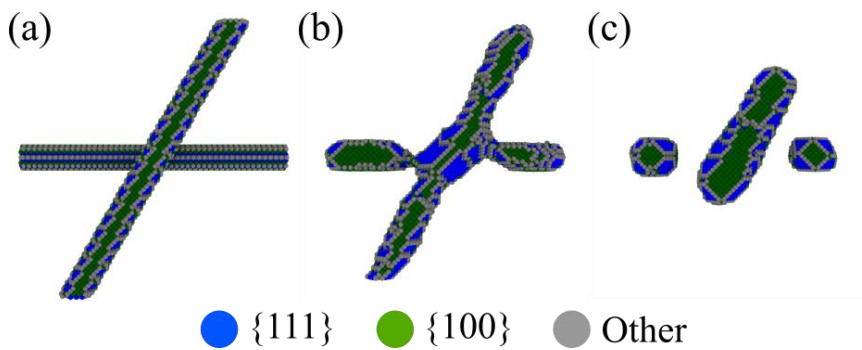

Figure 7. Breakup of a junction of non-periodic 1 nm radius wires at an oblique angle (1000 K). Initial configuration (a), intermediate state (b), and fragmented state (c). Available as movie 2 in supplementary materials.

Simulation of a less perfect system where the wires cross at an oblique angle (Figure 7) also results in the wires breaking at the junction, even though in this case the wires are much shorter (the system is non-periodic). The horizontal wire was <100>, as before.

In order to more closely approach the experimental five-fold twinned nanowires, we simulated crossing <110> nanowires. Figure 8 shows two periodic <110> wires breaking up around their point of contact. Because <110> nanowire breakup is much slower, the system had to be reduced in size to nanowire radius of 0.6 nm.

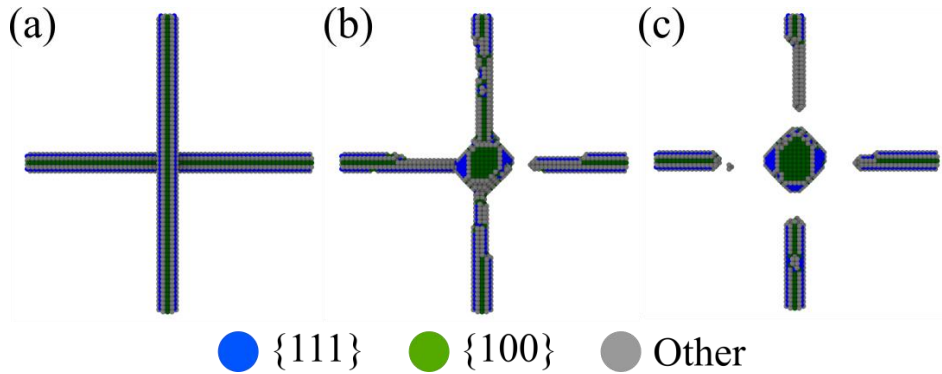

**Figure 8. Crossing 0.6 nm <110> wires that break around the junction (1000 K). Initial configuration (a), first detachment (b) and completely separated central cluster (c). Available as movie 3 in supplementary materials.**

We simulated an explicit array of nanowires. In a larger system, significantly more computational resources need to be spent to reach the same physical time. Even so, already at 5.4 ns we can see fragments forming at both junctions where two of the wires have separated (Figure 9).

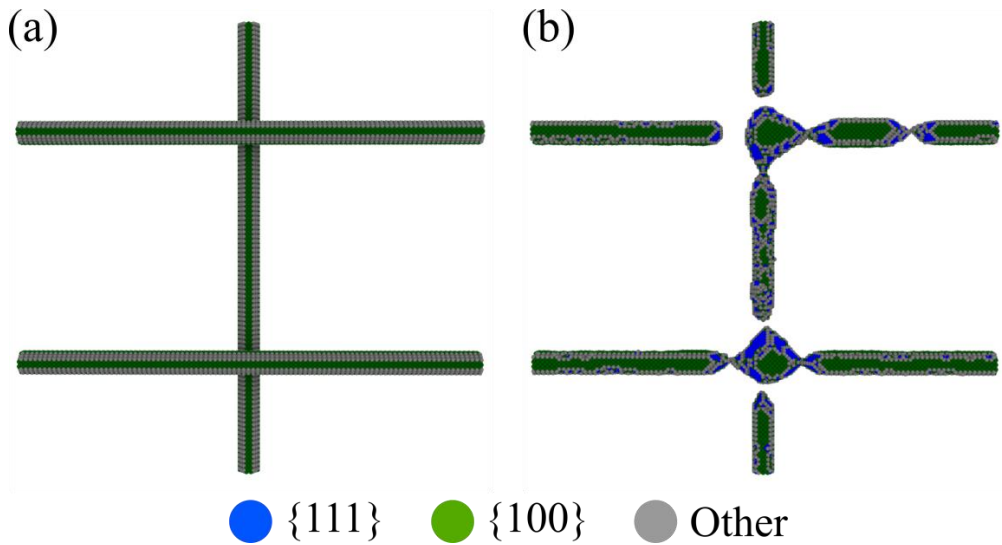

**Figure 9. Formation of fragments at two nearby junctions in an array of crossing nanowires. Initial configuration (a) and partially detached central fragments at 5.7 ns (b). Available as movie 4 in supplementary materials.**

## 4. Discussion

From the simulations we can see that nanowire breakup around junctions is driven by surface diffusion of atoms, similarly to the Rayleigh instability breakup. In case of a single nanowire, the size and positioning of the resulting clusters is random, though, the average size and separation can be predicted based on the initial nanowire radius. In case of a nanowire junction, however, a cluster always forms at the point where the wires intersect, and its size tends to be larger than the surrounding clusters due to the contribution of atoms from two wires instead of just one. Additionally, the breakup always starts at the junction because the intersecting nanowire surfaces act as defects, breaking the symmetry and encouraging atom diffusion.

In general, atoms have a higher probability to jump to sites with more neighbors. This explains the preference of atoms to accumulate at the wire crossing because the presence of two intersecting surfaces creates vacant sites with more neighbors than available on a single surface.

<100> FCC nanowires are unstable due to the shape memory effect [44] as they tend to rearrange their lattice to have the <110> direction along the wire. Such a rearrangement is outside the scope of fixed lattice KMC simulations. However, both <100> and <111> led to the same end result with fragmentation driven by surface energy minimization. Even though simulating the Rayleigh instability breakup of a single <110> wire proved impractical, placing two of them into contact in a reduced system resulted in a breakup process around the junction point, as seen in Figure 8. Therefore, it is reasonable to assume that the characterization of the junction breakups seen in <100> wire simulations is applicable also to the <110> case.

The wires used in experiment have a five-fold twinned structure, where the crystal direction along the wire is <110>. Because of the on-lattice nature of the KMC model, simulating such a structure is impossible. The closest structure, which we could use in the simulations to imitate the experiment, is to use a single crystal NW with the <110> direction along the wire. However, we can see similar structures forming both in experiments and in simulations irrespective of wire orientation, which indicates that the breakup process is driven by atom diffusion that is independent of the specific configuration.

Because of the thermally activated nature of the atomic diffusion, the timeframe of nanowire breakup is highly temperature dependent. Reducing the simulation temperature by just 20% resulted in a 30-fold increase in the time until the first detachment from the junction. In the experiment, treating nanowires at a temperature of 473 K (200 °C) for 10 minutes showed fragmentation almost exclusively at junctions only, which clearly shows the accelerating effect these sites have on nanowire fragmentation. Separate nanowires were significantly more stable at elevated temperatures.

The size of wires in simulations is necessarily much smaller than in experiments because of the large amount of computational resources required. However, as the KMC model does not include size effects, the result of atom diffusion is similar to the larger experimental systems. To further speed up the calculations, the temperature is also elevated compared to experiments. This is justified as we are still below melting temperature in its classical meaning, so that heating only accelerates processes that happen also at lower temperatures.

The breakup of nanowires due to Rayleigh instability has been observed for other FCC metals as well, and we have previously simulated this effect for Cu. Thus, the junction effect should behave in a similar manner for nanowires made of these metals. This is confirmed experimentally for the case of Ag, as seen in Figure 4. Here we note that the non-spherical shapes can be explained by rather low temperatures used in the experiments with Ag NWs (T=398 K; 125 °C), which resulted in only partial decomposition of the junction and an elongated central fragment. This is also observed for Au NWs in Figure 3 (a, b) for the temperatures ≤ 400 °C. At higher temperatures or longer thermal treatment times, all the NWs will break into fragments due to Rayleigh instability and the fragments will relax to spherical shapes as seen in Figure 3 (c) and (d). Because the temperature in KMC simulations is much higher (1000 K), fragments quickly become spherical, although elongated intermediate shapes can be seen in Figure 6 (d).

Because a cluster is always expected to form at a nanowire junction during annealing, we hypothesize that it is possible to fabricate regular arrays of nanodots by arranging nanowires in a grid and annealing them to induce the clusters to form at junctions. Furthermore, between the junctions, the nanowires will form nanodots with an average spacing $\lambda = 8.89 \cdot r$ given by the Rayleigh instability, where $r$ is the radius of the original wire [39,43]. Nanowires can be relatively easily arranged and aligned e.g. by dielectrophoresis [45]. The simulation with an array of wires (Figure 9) indicates that the clusters can be expected to form at the junction points with high reliability. When fragmentation is undesirable, a dense coating can be applied on nanowires to prevent diffusion of atoms as was found in present work.

## 5. Conclusions

We performed annealing experiments and corresponding Kinetic Monte Carlo simulations which show that two touching Au nanowires will break up in a specific manner where a cluster will form at the former junction of the nanowires. Annealing was conducted with Au and Ag nanowires at fixed temperatures 473, 673, 873 and 973 K (200, 400, 600 and 700 °C) during a time period of 10 minutes. In all cases, the junction breakup happened in a similar fashion for both metals. The experiments showed that junctions tended to break up even at lower temperatures when the wires themselves remained whole. We have developed a gold parametrization for the KMC code Kimocs which we have used to show that the breakup can be entirely explained by atom diffusion processes and the breakup of nanowires will always start at the junction. The point of nanowire contact acts as a preferential site for atomic diffusion due to the greater number of neighboring atoms present near surface intersections. The accumulation of atoms results in the formation of a cluster that is cut off from the nanowires. Thermal treatment significantly accelerates this process. We propose that nanowire junctions can be used to control the positioning of nanodots after thermal annealing of nanowires and that regular arrays of nanodots can be fabricated by aligning the nanowires in a grid.

## Acknowledgements

This work has been supported by Estonian Research council grants PUT 1372 and PUT 1689. V. Jansson was supported by Academy of Finland (Grant No. 285382) and Waldemar von Frenckells Stiftelse. E. Baibuz was supported by a CERN K-contract and the doctoral program MATRENA of the University of Helsinki. F. Djurabekova was supported by Academy of Finland (Grant No. 269696). The authors wish to acknowledge High Performance Computing Centre of University of Tartu and CSC – IT Center for Science, Finland, for computational resources. Authors are also grateful to Mikk Vahtrus for SEM images of thermally treated Ag nanowires.